\documentclass[12pt]{article}
\usepackage{amsmath,amssymb}
\usepackage{bm}
\usepackage{color}
\usepackage{graphicx}
\usepackage[english]{babel}
\usepackage{tikz-feynhand}
\usepackage{wrapfig}
\allowdisplaybreaks

\newcommand{\bea}{\begin{eqnarray}}
\newcommand{\eea}{\end{eqnarray}}

\newcommand{\bq}{\textbf{q}}

\newcommand{\nn}{\nonumber}

\def\be{\begin{equation}}
\def\ee{\end{equation}}

\def\bfx{{\bf x}}

\allowdisplaybreaks
\begin{document}
\title{
\begin{flushright}
\ \\*[-80pt] 
\begin{minipage}{0.2\linewidth}
\normalsize
$\quad$ HUPD-2101 \\*[5pt]
\end{minipage}
\end{flushright}
{\Large \bf 
Lepton family numbers and
non-relativistic \\ Majorana neutrinos
\footnote{This paper is based on a talk by T.M. in BSM-2021
online international conference.}
\\*[5pt]}}
\author{\normalsize
\centerline{Apriadi Salim Adam$^{1}$\footnote{E-mail address: apriadi.salim.adam@lipi.go.id},
Nicholas J. Benoit$^{2}$\footnote{E-mail address: d195016@hiroshima-u.ac.jp},
Yuta Kawamura$^{2}$\footnote{E-mail address: yuta-kawamura@hiroshima-u.ac.jp},
Yamato Matsuo$^{2}$\footnote{E-mail address: ya-matsuo@hiroshima-u.ac.jp}
} \\ \normalsize
\centerline{
Takuya Morozumi$^{3,4,}$\footnote{E-mail address: morozumi@hiroshima-u.ac.jp},
Yusuke Shimizu$^{3,4}$\footnote{E-mail address: yu-shimizu@hiroshima-u.ac.jp}, 
Yuya Tokunaga$^{5}$, and 
Naoya Toyota$^{2}$
}

\\*[10pt]
\centerline{
\begin{minipage}{\linewidth}
\begin{center}
$^1${\it \small
Research Center for Physics, Indonesian Institute of Sciences (LIPI),\\ Serpong PUSPIPTEK Area, Tangerang Selatan 15314, Indonesia}  \\*[5pt]
$^2${\it \small
Graduate School of Science, Hiroshima University, 
Higashi-Hiroshima 739-8526, Japan} \\*[5pt]
$^3${\it \small
Physics Program, Graduate School of Advanced Science and Engineering, \\
Hiroshima University, Higashi-Hiroshima 739-8526, Japan} \\*[5pt]
$^4${\it \small 
Core of Research for the Energetic Universe, Hiroshima University, \\
Higashi-Hiroshima 739-8526, Japan} \\*[5pt]
$^5${\it \small
 Hakozaki, Higashi-Ku, Fukuoka,~ 812-0053,~Japan
} \\*[5pt]
\end{center}
\end{minipage}}
\\*[50pt]}
\date{
\centerline{\small \bf Abstract}
\begin{minipage}{0.9\linewidth}
\medskip
\small
In this talk, we have reviewed the recent development on the time evolution of lepton family number carried by Majorana neutrinos \cite{Adam:2021qiq}.
This article focuses on the subtle points of the derivation of the lepton family numbers and their time evolution. We also show how the time evolution  is sensitive to
$m_{ee}$ and $m_{e\mu}$ components of the effective Majorana mass matrix by applying the formula to the two family case. The dependence on the Majorana phase is clarified
and the implication on CNB (cosmic neutrino background) is also discussed.
\end{minipage}
}
\begin{titlepage}
\maketitle
\thispagestyle{empty}
\end{titlepage}
\section{Introduction}
 The properties of neutrinos have not been fully understood yet. The following items are what we wish to address through our research.
One central question is that whether neutrinos are Majorana 
\cite{Majorana:1937vz} or Dirac \cite{Dirac:1928hu} particles and  how one can discriminate them. 
The nature of Majorana neutrinos is more pronounced at very low momentum $|\bq|<m_\nu$.
The temperature of CNB is expected to be as low as $T_{CNB}\simeq 2K \sim 2 \times 10^{-4}$ (eV).   In such low temperature, 
  even the neutrinos with the mass of order of $m_\nu=10^{-3}$ (eV) are non-relativistic.  

If the neutrinos are Majorana particle, the lepton number is not a conserved
quantity.  Traditionally,
the lepton number violation through the Majorana  mass \cite{Pontecorvo:1957qd} \cite{Bahcall:1978jn} \cite{Schechter:1980gk} \cite{Xing:2013ty}
and the lepton family number violation via neutrino oscillation \cite{Maki:1962mu}  have been  discussed separately.  The Majorana mass 
term generates the lepton number violation and it can lead to the neutrino-less double beta
decay
\cite{Furry:1939qr}.  The process occurs through the chirality flip of the Majorana neutrinos and its rate is suppressed as the neutrino's energy is about a few MeV for the typical
Nuclear reaction process.  
On the other hand, for the neutrino flavor oscillation, which is observed as the non-conservation of the charged lepton family numbers,  
the chirality flip is not relevant for the relativistic neutrinos and the neutrino oscillation formula has been derived without the effect.
It is known that in the relativistic limit, the common formula of the  neutrino flavor oscillation holds for  both Dirac neutrinos and  Majorana neutrinos and the CP asymmetry is insensitive to Majorana phases \cite{Doi:1980yb}, \cite{Schechter:1980gr}, \cite{Bilenky:1980cx}.
In the relativistic regime, one can not discriminate either Dirac or Majorana neutrino. 
By considering the fact that the temperature of CNB is comparable to the neutrinos mass with $10^{-4}$(eV),
it is desirable to formulate the lepton number violation and the lepton flavor violation in a single framework which is valid even in the non-relativistic region.
In our previous work \cite{Adam:2021qiq}, such framework has been developed for Majorana neutrino and the formula for time variation of lepton number which is valid for
the wide range of the momentum, i.e., from the relativistic case to non relativistic one, was derived.  In the framework, the lepton family number operators in the Heisenberg 
representation were derived by turning on the Majorana mass term in the Hamiltonian at $t=0$. This set up enables us to link the operators in the family basis to the operators
in the mass basis.  Then one can investigate the time evolution of the lepton family number operators under the presence of the Majorana mass terms.
See also \cite{Bittencourt:2020xen} \cite{Ge:2020aen} on the effect of the chirality flip of the Dirac neutrino and the implication on CNB.

In this paper, we review the paper by highlighting the subtle point of the derivation of
the time dependent lepton numbers. That is the 
continuity condition which relates the operator for chiral left-handed massless neutrino
with the definite family number $ e, \mu,$ and $\tau $  to the massive Majorana neutrino
with the definite mass.  We also apply the formula to the case with two family of neutrinos (electron and muon neutrinos).  The study of the simplified toy model is useful to understand how the time dependent lepton family numbers 
depend on the  Majorana phase, the absolute value of the neutrino masses, and the momentum of the neutrinos.
\section{The expectation value of the lepton number operator and how the lepton numbers are counted through the production and the detection of (anti-)neutrinos.}
To illustrate our set up, we start this section with a brief historical introduction from the view point of the lepton number violation. 
Bruno Pontecorvo \cite{Pontecorvo:1946mv} suggested the process which might be useful to detect the reactor neutrinos. The proposed process is given in the second line of  Eq.(\ref{eq:reaction}) .
In this idea, the neutrino from the reactor is supposed to be 
absorbed by $Cl_{17}^{34}$ and it subsequently decays into $Ar_{18}^{34}$.  Raymond Davis Jr. tried to observe the process and set an upper limit of the capture cross section \cite{Davis:1955bi}. If the reactor neutrino were  an electron neutrino, the absorption process can occur. However the reactor neutrino is an anti-electron neutrino and the absorption process with the electron production is  forbidden by the electron number conservation. With the analogy with $K^0$ and $\bar{K^0}$ mixing, Pontecorvo \cite{Pontecorvo:1957qd} reached to the idea of  $\bar{\nu}_e \rightarrow \nu_e $ oscillation ( neutrino-anti-neutrino oscillation ) and showed the process can occur if the neutrino is a Majorana particle as seen in Eq.(\ref{eq:reactionb})
\bea
n \rightarrow  p + e^-+&\bar{\nu}_e&  \mbox{reactor (anti-)neutrino}\nn \\
&&\searrow \nn \\
&&\quad \bar{\nu_e}+ Cl_{17}^{34} \rightarrow Ar_{18}^{34} +e^- .
\label{eq:reaction}
\eea
\bea
n \rightarrow  p + e^-+&\bar{\nu}_e& \quad  \mbox{reactor (anti-)neutrino}\nn \\
&&\searrow \nn \\
&&\quad \nu_e + Cl_{17}^{34} \rightarrow Ar_{18}^{34} +e^-.
\label{eq:reactionb}
\eea
Since the reactor anti-neutrino has the energy of the order of MeV and is relativistic, the anti-neutrino to neutrino transition is strongly suppressed by a factor of the neutrino mass over the energy.  Therefore the process such as Eq.(\ref{eq:reactionb}) could not be observed.

\begin{figure}[htpb]    
\begin{center}      
\begin{tikzpicture} 
\begin{feynhand}    
    \vertex [particle] (i3) at (-3,3) {$\pi^+$};
    \vertex [particle] (f1) at (0,4) {\small{$\langle \nu_e|L_{e}(0)|\nu_e \rangle  =1$}};
    \vertex [particle] (f2) at (1.5,3.5) {\small{$\qquad \langle \nu_e|L_{e}(t)|\nu_e \rangle=-1$}};
    \vertex [particle] (f3) at (3,3) {$\pi^0$};
     \vertex [particle] (f4) at (3,2) {$e^+$};
  \vertex [particle] (f5) at (0,3) ;
    \vertex [particle] (f6) at (1.5,3); 
\vertex [particle] (f7) at (0,2.7) ;
    \vertex [particle] (f8) at (1.5,2.7); 
    \vertex (w1) at (0.3,1.5) {$\nu_e$};
    \vertex (w2) at (0.8,1.2) {$\bar{\nu}_e$ };
    \vertex (w3) at (0,2) ;
    \vertex (w4) at (1.5,1);
    \vertex [particle] (e) at (3,1.5) {$e^{+}$};
     \vertex [particle] (an) at (3,0.5) {$W_{A-1}^{Z}$ };
    \vertex [particle] (an2) at (-2,0.5) {$ Y_{A}^{Z} $};
     \propag [antfer] (w3) to (f4);
    \propag [fermion] (i3) to (w3);
    \propag [fermion] (w3) to (f3);
    \propag [plain](f7) to [edge label=$t$] (f8); 
    \propag [maj](w3) to (w4)  [edge label];
    \propag [antfer] (w4) to (e);
    \propag [fermion] (an2) to (w4);
   \propag [fermion] (w4) to (an);
   \propag[plain](w3) to (f1);
 \propag[plain](w4) to (f2);
\end{feynhand}
\end{tikzpicture}
\caption{The production of electron neutrino and the detection of anti-neutrino through charged current interaction. This figure illustrates how the expectation value of the electron
number operator evolves with respect to time.  We have not observed neutrinos directly and  the whole process including the production and the detection corresponds to 
$\pi^+ + Y_A^Z \rightarrow \pi^0 + e^+ +( \nu_e \rightarrow  \bar{\nu}_e ) + Y_A^Z  \rightarrow \pi^0 + e^+ +  W_{A-1}^Z + e^+$.}
\label{fig:fig1}
\end{center}
\end{figure}
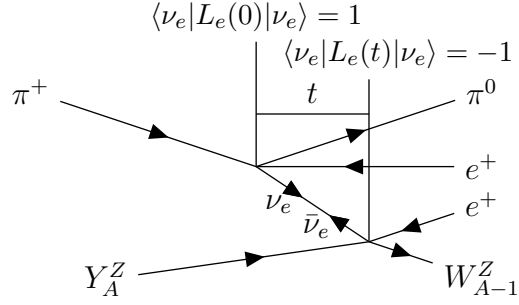

In Fig.\ref{fig:fig1}, we illustrate how the expectation value of electron number operator $\langle \nu_e|L_e(t)| \nu_e \rangle$ evolves with respect to time for the case that the electron neutrino is produced at $t=0$ and anti-neutrino is detected at time $t$. The neutrino's source is the charged pion decay $\pi^+ \rightarrow \pi^0 e^+ \nu_e$. The anti-electron neutrino
is detected with the inverse $\beta^+$ decay $ \bar{\nu}_e + Y_A^Z  \rightarrow e^+ +  W_{A-1}^Z$.
\section{Lepton family number carried by Majorana neutrinos}
Lepton family number ( electron number, muon number, and tauon lepton number) are assigned to SU(2) doublet of the three families.
Since Majorana neutrino is  a neutral particle, one can not assign the particle number associated with U(1) charge. The situation is similar to
the neutral scalar field which is described by a single component  Hermitian field.  The phase transformation on the neutral scalar is not allowed.
This is in contrast to the complex scalar field \cite{Morozumi:2017zyz} \cite{Hotta:2014ewa}.
 For the Dirac fermion which is similar to the complex scalar field, the fermion number can be defined with the time component of the vector current. However, if we do not restrict ourselves to the charge associated to the vector current, even for the Majorana field,
one can still define the family charge.
A Majorana field is written as a superposition of the left-handed chiral field and its charge conjugation. 
\bea
\psi_M=\nu_L +(\nu_L)^c.
\eea
Then one can define the lepton family charge of the left-handed neutrino as,
\bea
L=\int d^3 x :\bar{\nu_L}\gamma^0 \nu_L:.
\eea
Since only the left-handed neutrinos interact through weak interaction, we assign a lepton family number to the SU(2) doublet of each generation  where the charged leptons mass matrix is a real diagonal one.
\bea
\begin{pmatrix} 
\nu_{e_L} \\ e_L
\end{pmatrix},
\quad 
\begin{pmatrix} 
\nu_{\mu_L} \\ \mu_L
\end{pmatrix},
\quad
\begin{pmatrix} 
\nu_{\tau_L} \\ \tau_L
\end{pmatrix}.
\eea
The neutrino family eigenstates are not observed directly.  The family eigenstate of neutrino is tagged by the flavor of the charged lepton
which decays or is produced together with the neutrino in the charged weak interaction process. Therefore 
the lepton family number carried by the neutrino is written in terms of the left-handed chiral field,
\bea
L_\alpha= \int :\overline{\nu_{\alpha L}}\gamma^0 \nu_{\alpha L}: d^3x, \quad \alpha=e \sim \tau,
\label{eq:leptonfamily}
\eea
where the normal ordered product must be taken.
\section{The continuity condition and time evolution}
\begin{figure}[htbp]
\begin{center}
\includegraphics[width=0.45\linewidth]{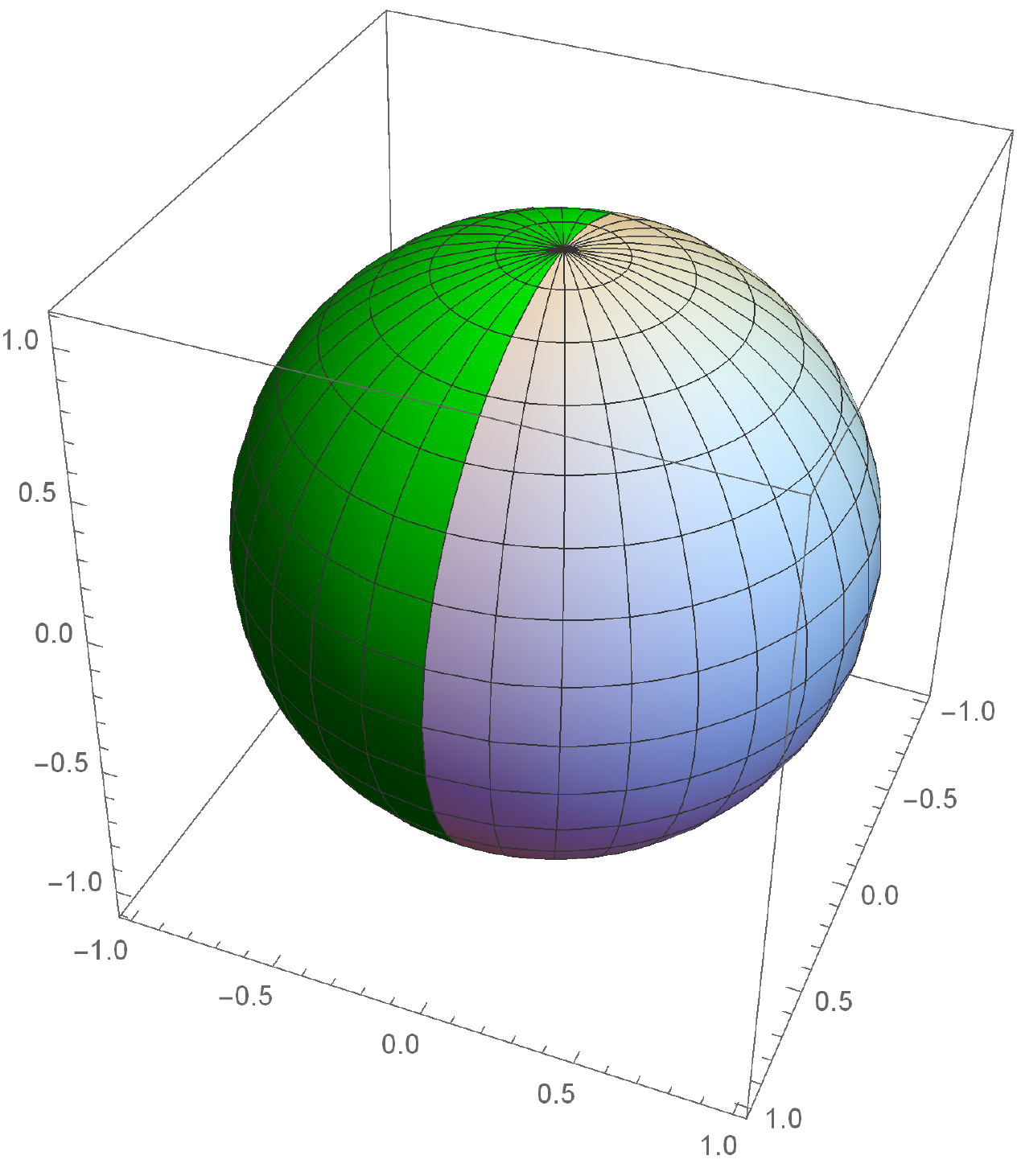}
\caption{The regions of the momentum direction ${\bf n}=\frac{\bf q}{|\bf q|}.$ The hemisphere colored with the light blue represents the region A and
the other hemisphere colored with the green corresponds to the region $\bar{A}$.}  
\label{fig:regions}
\end{center}
\end{figure}

We study the evolution of the lepton family number operators in the Heisenberg representation.
We have the following situation in mind.  When the weak charged current interaction produces the neutrino, we assume that
the neutrino is  produced as a $I_w=\frac{1}{2}$ partner of SU(2) doublet of the charged lepton.  
This corresponds to the following Lagrangian:
\bea
L=\overline{\nu_{L\alpha}} i \gamma_\mu \partial^\mu \nu_{L\alpha} -\theta(t) \frac{1}{2} \{m_{\alpha \beta} 
\overline{(\nu_{L\alpha})^c}(\nu_{L\beta})+h.c.\}.
\eea
Note that the mass term is turned at $t=0$ with the step function.
We introduce the mass eigenstates  $\psi_i$($i=1\sim 3$) so that the mass matrix given in the family basis is diagonalized by an unitary matrix V, 
\bea
&& (V^T)_{i \alpha} m_{\alpha \beta} V_{\beta j} = m_i \delta_{ij}, \nn \\
&& m_{\alpha \beta}= \sum_{i=1}^3 V^\ast_{\alpha i} m_i V^\ast_{\beta i},  
\eea
with 
\bea
\nu_{L \alpha}=V_{\alpha i} P_L \psi_{i}, \quad P_L=\frac{1-\gamma_5}{2}.
\label{eq:massbasis}
\eea
Below we explain the continuity condition. The equation of motion from the Lagrangian is,
\bea
\gamma_\mu \partial^\mu \nu_{L\alpha}=-i \theta(t)m^\ast_{\alpha \beta} (\nu_{L\beta})^c.
\eea
The time derivative of the field is given as, 
\bea
\frac{\partial}{\partial t}  \nu_{L\alpha}=-\gamma^0 \gamma^i \frac{\partial}{\partial x^i} \nu_{L\alpha}-i  \theta(t) m^\ast_{\alpha \beta}\gamma ^0 (\nu_{L\beta})^c. 
\eea
We integrate the both side of  the above equation with respect to time as,
\bea
&& \int_{-\epsilon}^\epsilon dt  \frac{\partial  \nu_{L\alpha}}{\partial t}=-\int_{-\epsilon}^{\epsilon} dt \gamma^0 \gamma^i \frac{\partial}{\partial x^i}
 \nu_{L\alpha}-i \int_0^{\epsilon} dt  \gamma^0  m^\ast_{\alpha \beta} (\nu_{L\beta})^c,
\eea
where the dependence on the space coordinate ${\bf x}$ is suppressed. By integrating the equation, one obtains, 
\bea
\nu_{L\alpha}(t=\epsilon) -\nu_{L\alpha}(t=-\epsilon)=-\int_{-\epsilon}^{\epsilon}  dt \gamma^0 \gamma^i \frac{\partial}{\partial x^i} \nu_{L\alpha}
 -i \int_0^{\epsilon} dt m^\ast_{\alpha \beta}   (\nu_{L\beta})^c.
\eea
At this stage, one can take the limit $\epsilon \rightarrow +0$ and obtains the continuity condition.
\bea
\nu_{\alpha L}(t=+0 , \bfx)=\nu_{\alpha L}(t=-0,\bfx) .
\eea
The left hand-side can be written with the mass basis introduced in Eq.(\ref{eq:massbasis}) through
\bea
\nu_{\alpha L}(t=+0,\bfx)=P_L V_{\alpha i} \psi_i(t=+0,\bfx) ,
\eea
where $P_L=\frac{1-\gamma_5}{2}$.
Then one can relate the operator for massless left-handed  one to chiral projection of the massive operator
which satisfies $(\psi_i)^c=\psi_i$. 
\bea
\nu_{\alpha L}(t=-0,\bfx)=P_L V_{\alpha i} \psi_i(t=+0,\bfx) .
\label{eq:continuity}
\eea
Note that the left-hand side of Eq.(\ref{eq:continuity}) is expanded with the annihilation operator of the neutrino ( $a_\alpha({\bf q})$ ) and the creation operator  for anti-neutrino  ( $b^\dagger_\alpha({\bf q})$ )   for $\alpha=(e,\mu,\tau) $ family. The right-hand side is written in terms of the superposition of the creation and annihilation operators ($a_{Mj}({\bf q},\lambda), j=1 \sim 3$)
for the Majorana fields with the definite masses ($m_j$) and the helicities ($\lambda=\pm$).   There is a subtle point of the momentum ${\bf q}$ carried by the neutrinos.
In the left-hand side of Eq.(\ref{eq:continuity}) , ${\bf q}=0$ is absent because zero momentum massless neutrino does not carry the energy. This in turn implies
the absence of  zero (${\bf q}=0$) mode for the left-hand side of Eq.(\ref{eq:continuity}). The continuity condition demands the absence of zero momentum mode for the massive Majorana operator
as well. This allows us to expand the massive Majorana field with the creation and annihilation operators with the definite helicity denoted as
$a^{(\dagger)}_{Mi}({\bf q}, \pm)$.   Since zero momentum
is excluded, one can split the entire momentum region of the non-zero momentum $\{ {\bf q}:  {\bf q}\ne 0 \}$ into two regions called $A$ and $\bar{A}$ as shown in Fig.2. 
By writing the direction 
of the momentum ${\bf n}=\frac{\bf q}{|{\bf q}|}$ with the  polar angles $\theta$ and $\phi$, the light blue-colored hemisphere in Fig.\ref{fig:regions} corresponds to the region where
the azimuthal angle $\phi$  lies within the range $[0,\pi)$. This region is named A region in \cite{Adam:2021qiq}. The other green-colored hemisphere corresponds to $\phi$ within 
$[\pi, 2\pi)$ and is named the region $\bar{A}$.   For 
the momentum ${\bf q}$ which lies in the A region,
the relations of the operators in family basis and those in mass basis are given as, 
\bea
\frac{a_\alpha({\bf q})}{\sqrt{2|{\bf q}|}}&=&\sum_{j=1}^3V_{\alpha j} \frac{\sqrt{N_j({\bf q})}}{2 E_j({\bf q})}
\left( a_{M j}({\bf q}, -)+ \frac{i  m_j}{ E_j({\bf q})+|{\bf q}|}  a^\dagger_{M j}({\bf -q}, -) \right), \label{eq:A}\\
\frac{a_\alpha({\bf -q})}{\sqrt{2|{\bf q}|}}&=&\sum_{j=1}^3 V_{\alpha j} \frac{\sqrt{N_j({\bf q})}}{2 E_j({\bf q})}
\left( a_{M j}({\bf -q}, -)- \frac{i  m_j}{ E_j({\bf q})+|{\bf q}|}  a^\dagger_{M j}({\bf q}, -) \right), \label{eq:Abar}
\eea
where 
$N_j({\bf q})=E_j({\bf q})+|{\bf q}|$ and $E_j({\bf q})=\sqrt{|{\bf q}|^2+m_j^2}$. The sign difference of the second terms of the right-hand side in 
 Eq.(\ref{eq:A}) and  Eq.(\ref{eq:Abar}) is consistent with the anti-commutation,
\bea
\{a_\alpha({\bf q}), a_\beta({\bf -q})\}=0.
\eea
The lepton family number operators defined in Eq.(\ref{eq:leptonfamily}) can be rewritten as,
\bea
L_\alpha(t=-0)=\int' \frac{d^3 q}{(2\pi)^3 2|{\bf q}|} \left(a^\dagger_\alpha({\bf q})  a_\alpha({\bf q})-b^\dagger_\alpha({\bf q})  b_\alpha({\bf q}) \right).
\label{eq:Lt0}
\eea
where $\int'$ implies that the zero momentum is excluded from the integration.
The relations in Eq.(\ref{eq:A}) and Eq.(\ref{eq:Abar}) lead to the time evolution of the creation and annihilation operators in the family basis. The explicit expression
for $a_\alpha({\bf q},t)$ and $b_\alpha({\bf q}, t)$ can be found in Eq.(30-33) of 
\cite{Adam:2021qiq}. The lepton family number at $t \ge 0$ is obtained by replacing $ a_\alpha({\bf q}) $and $b_\alpha({\bf q})$  in Eq.(\ref{eq:Lt0}) with  
$a_\alpha({\bf q},t) $and $b_\alpha({\bf q},t)$,
\bea
L_\alpha(t\ge 0)=\int' \frac{d^3 q}{(2\pi)^3 2|{\bf q}|} \left(a^\dagger_\alpha({\bf q},t)  a_\alpha({\bf q},t)-b^\dagger_\alpha({\bf q},t)  b_\alpha({\bf q},t)\right) .
\eea

 Next we apply the obtained formula to two families case, i.e.,  the electron and the muon families and show the expectation value of the electron and the muon number 
sandwiched with a state for a single electron neutrino with the momentum ${\bf q}$,
\bea
|{\bf q},e) \rangle =n_q a_e^\dagger({\bf q}) |0 \rangle,
\eea  
where the normalization constant $n_q$ is given by $\frac{1}{\sqrt{(2\pi)^3 2|{\bf q}| \delta^3({0})}}$.
The vacuum is defined so that it is annihilated by the annihilation operators
of family basis,
\bea
a_\alpha({\bf q})|0 \rangle=b_\alpha({\bf q})|0 \rangle=0.
\eea  
For the two family case, the mixing matrix is given as, 
\bea
V=\begin{pmatrix} c_{12} & s_{12}  e^{i \frac{\alpha_{21}}{2}}\\ 
-s_{12} & c_{12} e^{i \frac{\alpha_{21}}{2}}
\end{pmatrix},
\eea
where $\theta_{12}$ is a mixing angle, $s_{12}=\sin \theta_{12}$, and $c_{12}=\cos \theta_{12}$. $\alpha_{21}$ is a Majorana phase.
The expectation value of the electron number with respect to the electron neutrino state is,  
\bea
\langle\bq,e|L_{e}(t)|\bq,e\rangle&=& c_{12}^4
  \left(1-\frac{2m_1^2 \sin^2(E_1(\bq)t)}{E_1^2(\bq)}\right) 
+s_{12}^4\left(1-\frac{2m_2^2 \sin^2(E_2(\bq)t)}{E_2^2(\bq)}\right) \nn \\
&& +s_{12}^2 c_{12}^2 \left\{\left(1 +\frac{|\bq|^2-m_1 m_2 \cos(\alpha_{21})}
  {E_1(\bq)E_2(\bq)} \right) \cos(E_1(\bq)-E_2(\bq))t \right. \nn \\
&& + \left. \left(1 - \frac{|\bq |^2-m_1 m_2 \cos(\alpha_{21}) }
  {E_1(\bq)E_2(\bq)} \right) \cos (E_1(\bq)+E_2(\bq)) t \right\}. 
\label{eq:electron}
\eea
Similarly one obtains the muon number,
\begin{multline}
\langle\bq,e|L_{\mu}(t)|\bq,e\rangle = c_{12}^2 s_{12}^2
 \Bigl{[} \left(1-\frac{2m_1^2 \sin^2(E_1(\bq)t)}{E_1^2(\bq)}\right)
+ 
  \left(1-\frac{2m_2^2 \sin^2(E_2(\bq)t)}{E_2^2(\bq)}\right) \\
- \left( 1 + \frac{|\bq|^2-m_1 m_2 \cos(\alpha_{21})}{E_1(\bq)E_2(\bq)}\right)
  \cos(E_1(\bq)-E_2(\bq))t  
- \left(1 - \frac{|\bq|^2-m_1 m_2 \cos(\alpha_{21})}{E_1(\bq)E_2(\bq)}\right)
  \cos(E_1(\bq)+E_2(\bq))t  \Bigr{]}. 
\label{eq:muon}
\end{multline}
One can also obtain the total lepton number $\langle\bq,e|L(t)|\bq,e\rangle=\sum_{\alpha=e}^{\mu} \langle\bq,e|L_{\alpha} (t)|\bq,e\rangle$
and its upper and lower  bounds as,  
\bea
 \langle\bq,e|L(t)|\bq,e\rangle
&=& c_{12}^2 \left(1-\frac{2m_1^2 \sin^2(E_1(\bq)t)}{E_1^2(\bq)}\right)  
+s_{12}^2 \left(1-\frac{2m_2^2 \sin^2(E_2(\bq)t)}{E_2^2(\bq)}\right)  \nn \\
&=&\frac{|\bq|^2+m_1^2 \cos2 E_1(\bq) t}{E^2_1(\bq)} c_{12}^2+\frac{|\bq|^2+m_2^2 \cos2E_2(\bq)t}{E^2_2(\bq)} s_{12}^2, \nn \\
-1 & \leq & \frac{|\bq|^2-m_1^2 }{|\bq|^2+m_1^2} c_{12}^2+\frac{|\bq|^2-m_2^2 }{|\bq|^2+m_2^2} s_{12}^2  \leq \langle\bq,e|L(t)|\bq,e\rangle \leq 1.
\label{eq:total}
\eea
\begin{figure}[ht!]
\begin{tabular}{cc}
\includegraphics[width=0.45\linewidth]{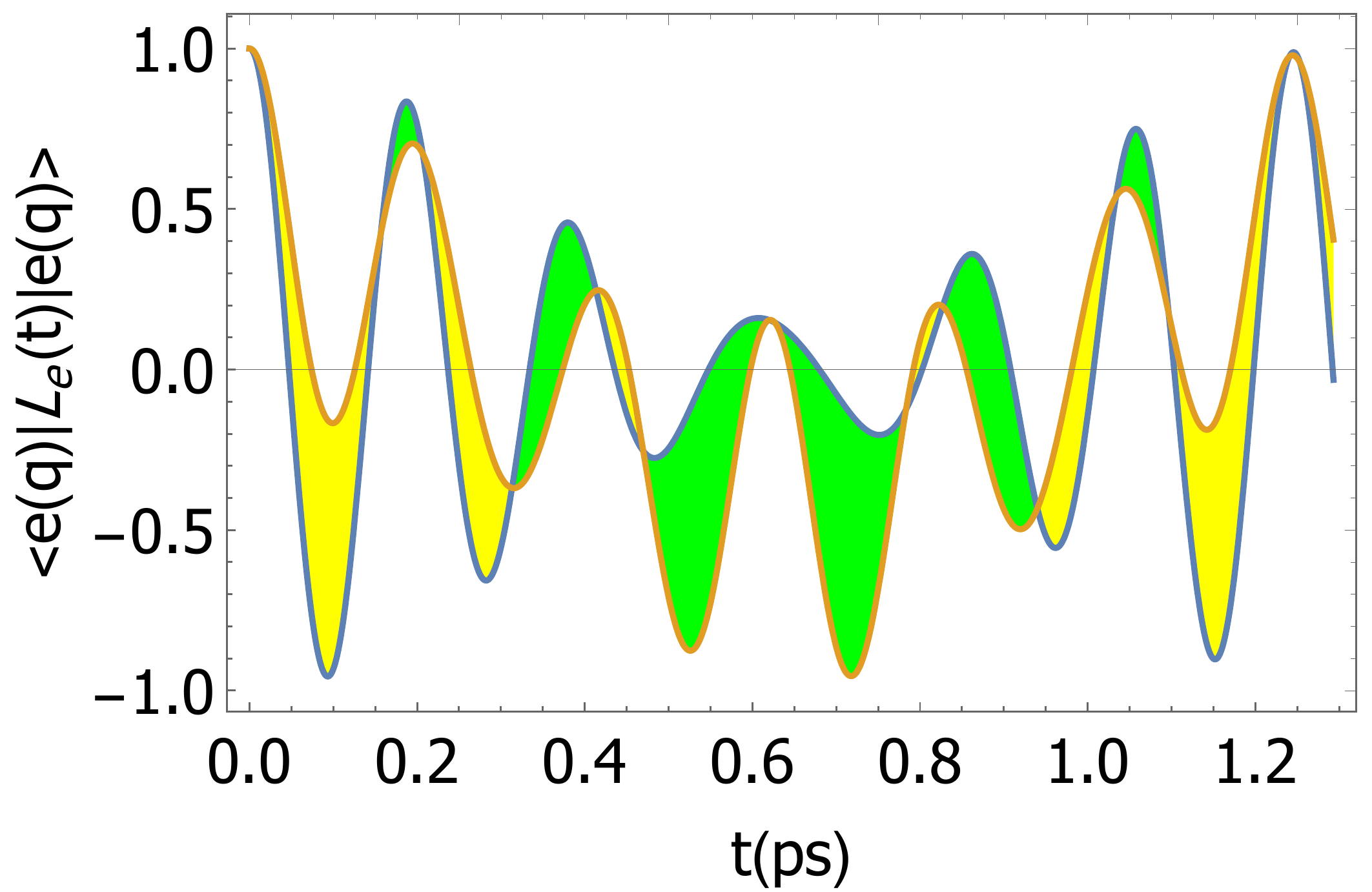}  
&
\includegraphics[width=0.45\linewidth]{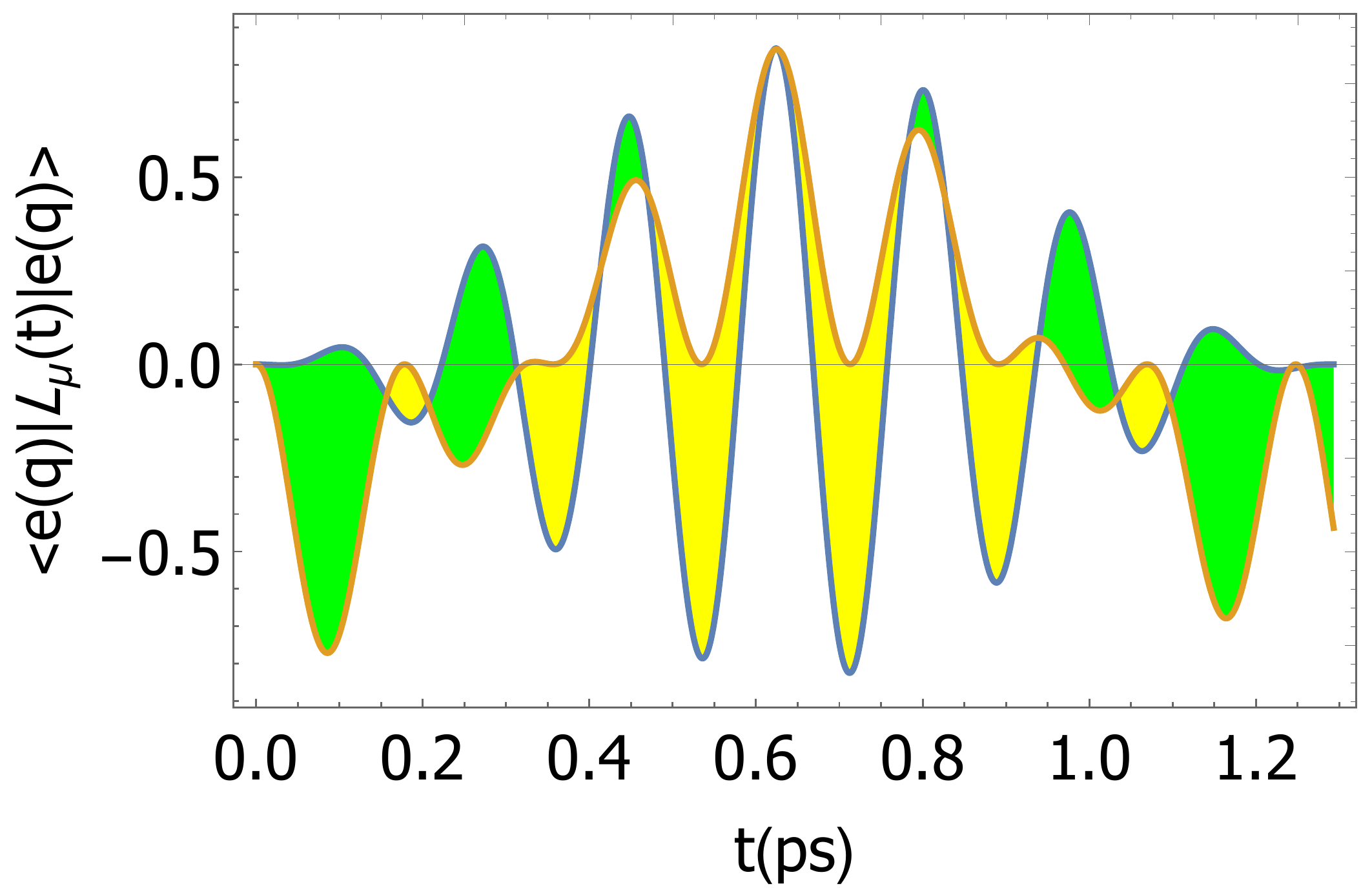}
 \\
electron number  &  muon number 
\end{tabular}
	\caption{
Time dependence of lepton family numbers $\langle e|L_{e}(t)|e \rangle$ 
and $\langle e|L_\mu(t)| e \rangle$ for two family model.
 The lightest neutrino mass is $0.01$ (eV) and the momentum of the neutrinos are $|\bq|=0.0002$ (eV).
The filled region corresponds to the change of a single  Majorana phase $\alpha_{21}$ 
within the range $[0,\pi]$. The mixing angle $\theta_{12}$ is chosen as 
$\sin\theta_{12}=0.551,  \cos\theta_{12}=0.834$ and $\Delta^2m_{21}=7.42 \times 10^{-5}$(eV$^2$). 
The blue colored lines show the case with the Majorana phase $\alpha_{21}=0$ while the light brown colored lines show the case with $\alpha_{21}=\pi$.
}
\label{fig:electronmuon}
\end{figure}
The lower bound of the total lepton number in Eq.(\ref{eq:total}) occurs at $|\bq|=0$ and it is $-1$.
As $|\bq|$ larger, it asymptotically reaches to +1.  
By expanding the expectation value of electron number in Eq.(\ref{eq:electron}) and muon number in Eq.(\ref{eq:muon}) around at  $t=0$, one obtains the following 
behavior, 
\bea
\langle\bq,e|L_{e}(t)|\bq,e\rangle
&\simeq& 1-2(|c_{12}^2 m_1^2 +s_{12}^2 m_2^2 e^{i \alpha_{21}}|^2) t^2 
-(m_1^2+m_2^2-2m_1m_2\cos(\alpha_{21}) ) s_{12}^2 c_{12}^2 t^2 \nn \\
&=&1-2|m_{ee}|^2 t^2-|m_{\mu e}|t^2, \nn \\
&=& 1-|m_{ee}|^2t^2-(m^\dagger m)_{ee}t^2. \\
\langle\bq,e|L_{\mu}(t)|\bq,e\rangle
&\simeq& -c_{12}^2 s_{12}^2 ( m_1^2+m_2^2-2 m_1m_2\cos\alpha_{21}) t^2,\nn \\
&=&-|m_{\mu e}|^2 t^2.
\eea
where $|m_{ee}|^2$ ,$|m_{\mu e}|^2$ and $(m^\dagger m)_{ee}$ are respectively given as,
\bea
&& |m_{ee}|^2=|m_1 c_{12}^2+m_2
 s_{12}^2 e^{-i \alpha_{21}}|^2, \label{eq:mee} \\
&& |m_{\mu e}|^2=|s_{12} c_{12}(m_2 e^{-i \alpha_{21}}-m_1)|^2 ,\label{eq:mmue} \\
&& (m^\dagger m)_{ee}=(V^\dagger m^2 V)_{ee}=m_1^2 c_{12}^2 +m_2^2 s_{12}^2.
\eea
In Fig.\ref{fig:electronmuon} , 
we show the expectation values for electron number and muon
number for two different values of Majorana phases $\alpha_{21}=0$ and $\pi $. When Majorana phase is chosen as $\alpha_{21}=0$, $|m_{ee}|$ is the maximum 
and $|m_{\mu e}|$ is the minimum as can be seen from Eq.(\ref{eq:mee}) and Eq.(\ref{eq:mmue}). 
For $\alpha_{21}=\pi$, $|m_{ee}|$ is the minimum  and $|m_{\mu e}|$ is the maximum.
The behavior of the electron number and muon number in Fig.\ref{fig:electronmuon} around $t=0$  are consistent with the observation described above.
The blue colored lines  in Fig.\ref{fig:electronmuon}  show the case with the Majorana phase $\alpha_{21}=0$ while the light brown colored lines show the case with $\alpha_{21}=\pi$. In the latter case, the muon number sharply decreases around at $t=0$ and it implies large $\nu_e \to \bar{\nu}_\mu$ conversion rate. For the case $\alpha_{21}=0$, the 
electron number sharply decreases because of large $|m_{ee}|$. 
In Fig.\ref{fig:qdependence}, we have shown the expectation values at specific time $t=0.66$ (ps). The figure shows the dependence on the Majorana phase is more significant as we
lower the momentum. 
\begin{figure}[ht!]
\begin{center}
\includegraphics[width=0.6\linewidth]{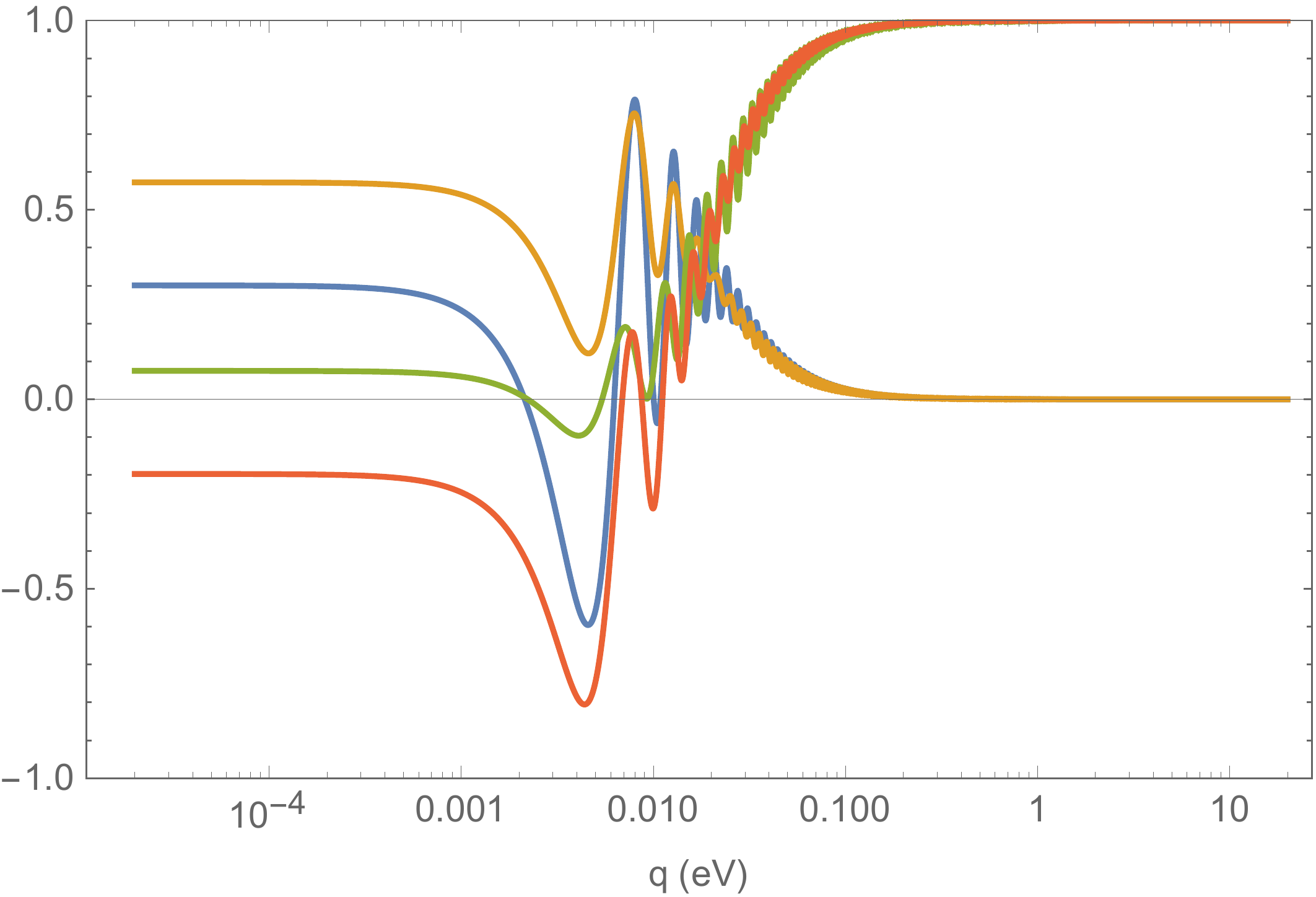} 
\caption{The momentum $q=|{\bf q}|$ dependence of the expectation values at $t=0.66$(ps). The yellow colored curve and the blue colored curve show the muon numbers, i.e.,  $\langle e|L_{\mu}(t=0.66)|e \rangle$ for $\alpha_{21}=\pi$ and $\alpha_{21}=0$, respectively.  The green colored curve and the orange colored curve show the electron number, i.e., 
$\langle e|L_{e}(t=0.66)|e \rangle$ for  $\alpha_{21}=0$ and $\alpha_{21}=\pi$, respectively.
 }
\end{center}.
\label{fig:qdependence}
\end{figure}
\section{Summary and Implication on CNB}
 With the lepton family number, the neutrino flavor transition and the neutrino and anti-neutrino  transition effects is included in a single framework and
the formula for time evolution of lepton numbers that is valid for the broad range of the momentum was obtained \cite{Adam:2021qiq}. 
We reviewed this approach by focusing on 
the definition of the lepton family numbers, the several subtle points of the derivation such as the continuity condition and the momentum regions. 
We apply the formula to the two generation case of neutrinos and show how the expectation values depend on the Majorana phase and their momentum.
The effect of Majorana mass term including the Majorana phase is more  significant at the low momentum region $q<m_{\nu}$.

The implication of the present work on the CNB can be summarized as follows. In the early universe, neutrinos' momentum is large and the lepton family numbers are approximately conserved. While the universe is cooled down,
the momentum carried by the neutrinos is red-shifted and they become non-relativistic. If the neutrinos are Majorana particles, the effects of Majorana mass term is turned on, and the lepton number may oscillate and alternate its sign . These effects  strongly depend on the Majorana phases, neutrino masses as we have seen.
See also \cite{Bittencourt:2020xen} \cite{Ge:2020aen} on the effect of the chirality flip of the Dirac neutrino and the implication on CNB.

\vspace{1cm}

\noindent
{\bf Acknowledgement}\\
We would like to thank the organizers of the BSM-2021. The work of T.M. is supported by Japan Society for the Promotion of Science (JSPS) KAKENHI Grant Number JP17K05418.


\begin{thebibliography}{00}
\bibitem{Adam:2021qiq}
A.~S.~Adam, N.~J.~Benoit, Y.~Kawamura, Y.~Matsuo, T.~Morozumi, Y.~Shimizu, Y.~Tokunaga and N.~Toyota,
``Time Evolution of Lepton Number Carried by Majorana Neutrinos,''doi.org/10.1093/ptep/ptab025,
[arXiv:2101.07751 [hep-ph]].
\bibitem{Majorana:1937vz}
  E.~Majorana,
  Nuovo Cim.\  {\bf 14} (1937) 171.
\bibitem{Dirac:1928hu}
P.~A.~M.~Dirac,
Proc. Roy. Soc. Lond. A \textbf{117}, 610-624 (1928)
doi:10.1098/rspa.1928.0023.
\bibitem{Pontecorvo:1957qd}
  B.~Pontecorvo,
  Sov.\ Phys.\ JETP {\bf 7}, 172 (1958)
  [Zh.\ Eksp.\ Teor.\ Fiz.\  {\bf 34}, 247 (1957)].

\bibitem{Bahcall:1978jn}
  J.~N.~Bahcall and H.~Primakoff,
  Phys.\ Rev.\ D {\bf 18} (1978) 3463.
\bibitem{Schechter:1980gk}
  J.~Schechter and J.~W.~F.~Valle,
  Phys.\ Rev.\ D {\bf 23} (1981) 1666.

\bibitem{Xing:2013ty}
  Z.~z.~Xing,
  Phys. Rev. D \textbf{87}, no.5, 053019 (2013).



\bibitem{Maki:1962mu} 
  Z.~Maki, M.~Nakagawa and S.~Sakata,
  Prog.\ Theor.\ Phys.\  {\bf 28}, 870 (1962).

\bibitem{Furry:1939qr}
  W.~H.~Furry,
  Phys.\ Rev.\  {\bf 56} (1939) 1184.

\bibitem{Doi:1980yb} 
  M.~Doi, T.~Kotani, H.~Nishiura, K.~Okuda and E.~Takasugi,
  Phys.\ Lett.\  {\bf 102B}, 323 (1981).

\bibitem{Schechter:1980gr}
J.~Schechter and J.~W.~F.~Valle,
Phys. Rev. D \textbf{22}, 2227 (1980)
doi:10.1103/PhysRevD.22.2227.

\bibitem{Bilenky:1980cx}
S.~M.~Bilenky, J.~Hosek and S.~T.~Petcov,
Phys. Lett. B \textbf{94}, 495-498 (1980)
doi:10.1016/0370-2693(80)90927-2.

\bibitem{Bittencourt:2020xen}
V.~A.~S.~V.~Bittencourt, A.~E.~Bernardini and M.~Blasone,
[arXiv:2009.00084 [hep-ph]].

\bibitem{Ge:2020aen}
S.~F.~Ge and P.~Pasquini,
Phys. Lett. B \textbf{811}, 135961 (2020)
doi:10.1016/j.physletb.2020.135961
[arXiv:2009.01684 [hep-ph]].

\bibitem{Pontecorvo:1946mv}
B.~Pontecorvo,
Camb. Monogr. Part. Phys. Nucl. Phys. Cosmol. \textbf{1}, 25-31 (1991)
PD-205.

\bibitem{Davis:1955bi}
R.~Davis, Jr.,
Phys. Rev. \textbf{97}, 766-769 (1955)
doi:10.1103/PhysRev.97.766.
\bibitem{Morozumi:2017zyz}
  T.~Morozumi, K.~I.~Nagao, A.~S.~Adam and H.~Takata,
  Adv. High Energy Phys. \textbf{2019}, 6825104 (2019).
\bibitem{Hotta:2014ewa}
  R.~Hotta, T.~Morozumi and H.~Takata,
  Phys. Rev. D \textbf{90}, no.1, 016008 (2014).
\end{thebibliography}
\end{document}